\documentclass{INTERSPEECH2023}
\interspeechcameraready 

\usepackage[hyphens,spaces,obeyspaces]{url}
\usepackage[colorlinks]{hyperref}

\usepackage{enumitem}
\usepackage{bm}
\usepackage{amssymb}
\usepackage{amsmath}
\usepackage{graphicx}
\usepackage{algpseudocode}
\usepackage{enumerate}
\usepackage{booktabs}
\usepackage{makecell}
\usepackage{threeparttable}
\usepackage{tabularx} 
\usepackage{multirow}
\usepackage{ragged2e}
\usepackage{xcolor}
\usepackage[normalem]{ulem}

\usepackage{cite}

\title{SVDD Challenge 2024: A Singing Voice Deepfake Detection Challenge Evaluation Plan}
\name{You Zhang$^{*1}$, Yongyi Zang$^{*1}$, Jiatong Shi$^{*2}$, Ryuichi Yamamoto$^{*3}$, Jionghao Han$^2$, Yuxun Tang$^4$,  \ Tomoki Toda$^3$, Zhiyao Duan$^1$\thanks{* These authors contributed equally. Version 1.2. Updated: \today. Correspondence addressed to \texttt{you.zhang@rochester.edu}}}

\address{$^1$University of Rochester, Rochester, NY, USA 
$^2$Carnegie Mellon University, Pittsburgh, PA, USA 
$^3$Nagoya University, Nagoya, Japan
$^4$Renmin University of China, Beijing, China 
}
\email{svddchallenge@gmail.com \\ \url{https://challenge.singfake.org}}

\begin{document}

\maketitle

\begin{abstract}
The rapid advancement of AI-generated singing voices, which now closely mimic natural human singing and align seamlessly with musical scores, has led to heightened concerns for artists and the music industry. Unlike spoken voice, singing voice presents unique challenges due to its musical nature and the presence of strong background music, making singing voice deepfake detection (SVDD) a specialized field requiring focused attention. To promote SVDD research, we recently proposed the ``SVDD Challenge,'' the very first research challenge focusing on SVDD for lab-controlled and in-the-wild bonafide and deepfake singing voice recordings. The challenge will be held in conjunction with the 2024 IEEE Spoken Language Technology Workshop (SLT 2024).
\end{abstract}


\section{Introduction}
The development of advanced singing voice synthesis techniques has led to a significant milestone in AI-generated content, where singing voices sound remarkably natural and align seamlessly with music scores. These synthesized voices can now emulate the vocal characteristics of any singer with minimal training data. While this technological advancement is impressive, it has sparked widespread concerns among artists, record labels, and publishing houses~\cite{collins2024avoiding}. The potential for unauthorized synthetic reproductions that mimic well-known singers poses a real threat to original artists' commercial value and intellectual property rights, igniting urgent calls for efficient and accurate methods to detect these deepfake singing voices.

In response to these concerns, our prior research~\cite{zang23svdd} has laid the groundwork for the emerging field of SVDD. We introduced the SingFake dataset, a comprehensive collection of authentic and deepfake song clips featuring a variety of languages and singers. Our findings revealed a critical insight: conventional speech countermeasure (CM) systems, when trained on standard speech, experience significant performance degradation when tested on singing voices. Conversely, retraining these systems specifically on singing voice data resulted in notable improvements. Our prior evaluations also highlighted several challenges, including dealing with unseen singers, various communication codecs, diverse languages and musical contexts, and interference from accompaniment tracks. This highlights the distinct nature of singing voice deepfakes and the necessity for specialized SVDD systems. 

To advance the field of singing voice deepfake detection, we introduce the SVDD challenge, the inaugural research initiative specifically dedicated to exploring SVDD. This challenge targets both controlled and in-the-wild settings, aiming to distinguish bonafide and AI-generated singing voice recordings.

\section{Challenge objectives}

The SVDD challenge aims to bring together the academic and industrial communities to develop innovative and effective techniques for detecting deepfake singing voices. We hope this challenge will advance our understanding of the specific traits of singing voice deepfakes and contribute to the broader field of multimedia deepfake detection.

\section{Challenge setups}
In the context of singing voice deepfakes, a common practice is to present these artificial creations alongside background music, as observed in our SingFake~\cite{zang23svdd} project. This approach, while practical for simulating authentic song presentations, introduces a significant challenge: the separation of vocals from music may create artifacts that can obscure the differences between bonafide and deepfake vocals. To investigate this issue thoroughly, the SVDD challenge is structured into two distinct tracks: the \textit{controlled} and the \textit{in-the-wild} settings. The WildSVDD track follows the same approach as our SingFake project~\cite{zang23svdd}, dealing with deepfakes as they typically appear in online media, complete with background music. In contrast, the CtrSVDD track exclusively uses clean, unaccompanied vocals provided by our data contributors, thereby minimizing the interference of voice separation algorithms. This two-track approach lets participants tackle the challenges of identifying deepfake singing voices under different and realistic conditions. This section describes our dataset curation for setting up the two tracks.

\subsection{CtrSVDD: Controlled singing voice deepfake detection}


For the CtrSVDD track, we first source bonafide datasets from existing open-source singing recordings. These include Mandarin singing datasets: Opencpop~\cite{wang2022opencpop}, M4Singer~\cite{zhang2022m4singer}, Kising~\cite{shi2024singing}, Official ACE-Studio release, and Japanese singing datasets: Ofuton-P\footnote{\url{https://sites.google.com/view/oftn-utagoedb/\%E3\%83\%9B\%E3\%83\%BC\%E3\%83\%A0}}, Oniku Kurumi\footnote{\url{https://onikuru.info/db-download/}}, Kiritan~\cite{ogawa2021tohoku}, and JVS-MuSiC~\cite{tamaru2020jvs}. We perform segmentation to divide the songs into vocal clips.

We then generate deepfake singing vocals with 14 existing singing voice synthesis~(SVS) and singing voice conversion~(SVC) systems from these bonafide vocals. For SVS, we employ ESPnet-Muskits~\cite{shi2022muskits}, NNSVS~\cite{yamamoto2023nnsvs}, DiffSinger~\cite{liu2022diffsinger}, and ACESinger\footnote{\url{https://acestudio.ai/}}. For SVC, we apply the NU-SVC~\cite{yamamoto2023comparative} and variants of So-VITS-SVC\footnote{
\url{https://github.com/HANJionghao/so-vits-svc2}}. When dividing the generated deepfakes into train, development, and evaluation splits, we follow the approach used in the speech deepfake detection benchmark ASVspoof2019. We employ the same set of deepfake generation algorithms (A01-A08) for both the training and development sets, while using a different set of deepfakes (A09-A14) for the evaluation partition. 
We plan to release more details on deepfake systems in early June.


\subsection{~\hspace{-10pt} WildSVDD: In-the-wild singing voice deepfake detection}

We have continued to gather data annotations from social media platforms following a method similar to the SingFake project~\cite{zang23svdd}. The WildSVDD dataset has been expanded to approximately double the original size of SingFake, now featuring 97 singers with 3223 songs. The annotators, who are familiar with the singers they cover, manually verified the user-specified labels during the annotation process to ensure accuracy, especially in cases where the singer(s) did not actually perform certain songs. 
We cross-check the annotations against song titles and descriptions, and manually review any discrepancies for further verification.
We have verified the accessibility of all URLs in the dataset as of March 28th and removed any that were inaccessible. The WildSVDD dataset now includes Korean singers, making Korean the third most represented language in the dataset. To help track changes between the SingFake and WildSVDD datasets, we have added a "SingFake\_Set" column that indicates the original partition of an annotation in the SingFake dataset. Annotations that lack a value in this column are new additions to the WildSVDD dataset.

Due to potential copyright issues, we are currently only releasing the annotations. Consequently, participants might acquire slightly different media files that correspond to the same annotations, depending on the specifics of their download process. Due to this variability, self-reported metrics from participants can, at best, be used as a rough reference and cannot be directly used to compare systems.
As such, we encourage participants to report the success rate of URL downloads per partition and, if possible, the actual files used for training and testing. This transparency will allow researchers to make fairer comparisons. Additionally, participants are encouraged to describe their model designs clearly and open-source their model implementations to facilitate the reproduction of results with the WildSVDD dataset.

\section{Evaluation metrics}

\begin{figure}[t]
\centering
\includegraphics[width=0.4\textwidth]{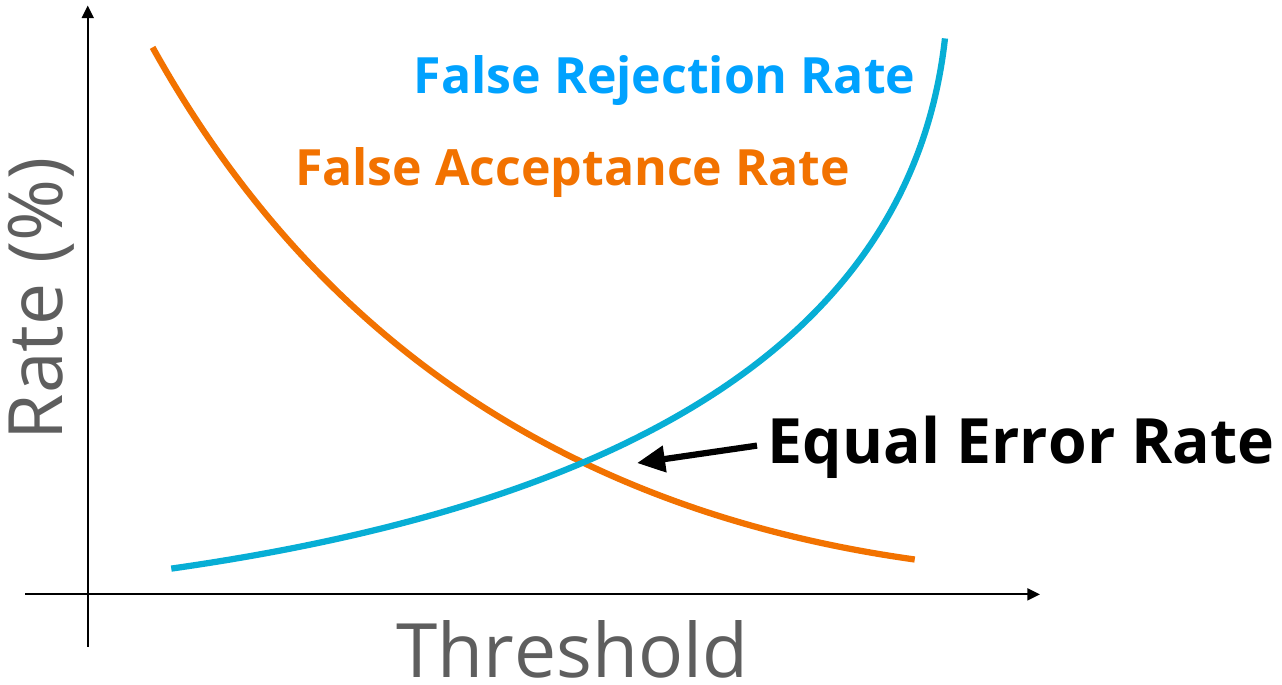}
\caption{Illustration of Equal Error Rate (EER).}
\label{fig:eer}
\end{figure}

We use Equal Error Rate (EER) to evaluate the SVDD system performance. 
We expect each SVDD system submitted by participants to generate a score TXT file, which contains scores for every segmented clip. These scores reflect the system's confidence in whether the vocal or song clip originates from a bona fide singer or resembles a real singer. A higher score indicates greater confidence that the clip is from a real singer.
In practical usage, people may set a threshold to determine the binary output of bonafide or deepfake. With the threshold higher, the false acceptance rate will become lower, and the false rejection rate will become higher. The EER is achieved when these two are equal, as illustrated in Figure~\ref{fig:eer}. The EER is more suitable to evaluate the system’s performance than accuracy since it is not dependent on the threshold. The lower the EER, the better the system distinguishes bonafide and deepfake singing voices.

\section{Protocols}
\subsection{CtrSVDD track}
The CtrSVDD data is released under a CC-BY-NC-ND 4.0 license, aligned with the sourcing corpora.
We have released the training and development set on Zenodo\footnote{\url{https://zenodo.org/records/10467648}} and other relevant scripts on GitHub\footnote{\url{https://github.com/SVDDChallenge/CtrSVDD_Utils}}. 
Please be aware that the training and development set available on Zenodo is incomplete because of licensing issues. To fully generate the dataset, first download all the remaining bonafide datasets on your own by agreeing to their terms and conditions, and then follow the detailed instructions provided in the GitHub repository. Participants can refer to the statistics in Table~\ref{table: CtrSVDD} as a guide to verify the completion of their downloads and generation.

For evaluation, we have released the test set on Zenodo\footnote{\url{https://zenodo.org/records/10742049}} with undisclosed labels at a later date and ask teams to score each song vocal clip. There are, in total, 48 speakers with 92,769 clips. Using the submitted scores, we will calculate and rank participant systems using EER. 
For submission guidelines, please refer to Section~\ref{sec: submit}.



\begin{table}[t]
\centering
\caption{Summary of the training and development partition of the CtrSVDD dataset.}
\label{table: CtrSVDD}
\resizebox{\columnwidth}{!}{
\begin{tabular}{@{}lllll@{}}
\hline\hline
\multirow{2}{*}{Partition} & \multirow{2}{*}{\begin{tabular}[c]{@{}c@{}}\# Speakers\end{tabular}}  
& Bonafide     
& \multicolumn{2}{c}{Deepfake} \\
\cmidrule(l){3-5}
& &\# Utts & \# Utts   
& Attack Types     \\
\midrule
Train   & 59      & 12,169  & 72,235 & A01$\sim$A08     \\
Dev & 55     & ~~6,547  & 37,078 & A01$\sim$A08     \\
\hline\hline
\end{tabular}}
\end{table}

\subsection{WildSVDD track} 



The WildSVDD track data has been released on Zenodo\footnote{\url{https://zenodo.org/records/10893604}} under a CC-BY 4.0 license. We supply the training and test partitions, allowing participants the flexibility to carve out a validation set from the training data for model development. We provide labels of SingFake~\cite{zang23svdd} partitions for annotations that appeared in the SingFake dataset for easy comparison with previous systems.
The test set is divided into parts A and B, with part B considered more challenging due to its inclusion of unseen musical contexts as T04 in SingFake~\cite{zang23svdd}.

We recommend that participants further segment the songs into clips using our tool available in the SingFake GitHub repository\footnote{\url{https://github.com/yongyizang/SingFake}}. Evaluations should be conducted at the segment level rather than at the song level.
We will adopt the self-reported EER and will not accurately rank the results. We encourage the participants to submit the score files listing the URLs, segment index, and the corresponding scores output from their systems.


\section{Rules for developing SVDD systems}


Participants are welcome to use any publicly available datasets for training in addition to the CtrSVDD we provide, but of course, exclude any datasets used in our test set. Specifically, for the CtrSVDD track, participants must NOT use M4Singer, KiSing, any open-sourced deepfake models based on M4Singer and/or KiSing, or the commercial software ACE Studio\footnote{\url{https://ace-studio.timedomain.cn/}}.

We refer the participants to the list of available datasets at the end of this section. However, participants must document any additional data sources used in their system descriptions. If there are any public data sources not listed but you would like to use for training, please inform the organizers so that we can share this information among participants. We will maintain and update the list of data sources until the registration deadline.

If the participants are willing to generate new training data from our released data and other public datasets for the CtrSVDD track, they can request permission to use such data under the condition that this new dataset will be published to other participants. We have set a deadline for this request. Please refer to Section~\ref{sec: dates} for details.

The use of publicly available pre-trained models is also permitted. Participants should specify the exact version of the pre-trained models used and provide a link to the pre-trained embedding used in the system description. 

Any private data or private pre-trained models are strictly forbidden to use. 


Participants should not add additional annotations to the WildSVDD track for training. 
Please contact the organizers if you are interested in contributing more annotations for future research.

Below we provide a list of known data sources as a reference. This list applies to both CtrSVDD and WildSVDD tracks.

\begin{itemize}

    \item  Speech Anti-Spoofing datasets: ASVspoof 2019~\cite{wang2020asvspoof}, ASVspoof 2021~\cite{liu2023asvspoof}, In-the-wild~\cite{muller2022does}, WaveFake~\cite{frank2021wavefake}

    \item  Speech Synthesis datasets: LJSpeech~\cite{ito2017ljspeech}, VCTK~\cite{veaux2017cstr}, LibriTTS~\cite{zen2019libritts}, Hi-Fi TTS~\cite{bakhturina2021hi}, LibriSpeech~\cite{panayotov2015librispeech}, CommonVoice~\cite{ardila2020common}, LibriLight~\cite{kahn2020libri}, 
    
    \item Singing Voice Synthesis datasets: NUS-48E~\cite{duan2013nus}, OpenSinger~\cite{huang2021multi}, CSD~\cite{choi2020children}, VocalSet~\cite{wilkins2018vocalset}, Ameboshi\_ciphyer\_utagoe\_db\footnote{\url{https://parapluie2c56m.wixsite.com/mysite}}, itako\_singing\footnote{\url{https://github.com/mmorise/itako_singing}}, JSUT\footnote{\url{https://sites.google.com/site/shinnosuketakamichi/publication/jsut-song}}, Namine\_ritsu\_utagoe\_db\footnote{\url{https://drive.google.com/drive/folders/1XA2cm3UyRpAk_BJb1LTytOWrhjsZKbSN}}, Natsume\footnote{\url{https://bowlroll.net/file/224647}}, NIT\_song070\footnote{\url{http://hts.sp.nitech.ac.jp/}}, No7\_singing\footnote{\url{https://github.com/mmorise/no7_singing}}, PJS~\cite{koguchi2020pjs}, PopCS~\cite{liu2022diffsinger}, Dsing~\cite{dabike2019automatic}, SingStyle111~\cite{dai2023singstyle111}

    \item Audio-Visual Singing Voice datasets: URSing~\cite{Li-2021}, A cappella~\cite{montesinos2021cappella}

    \item  Singing Voice DeepFake Detection datasets: SingFake \cite{zang23svdd}, FSD~\cite{xie2023fsd}.

\end{itemize}

\section{Baselines}

\begin{figure}[t]
\centering
\includegraphics[width=0.4\textwidth]{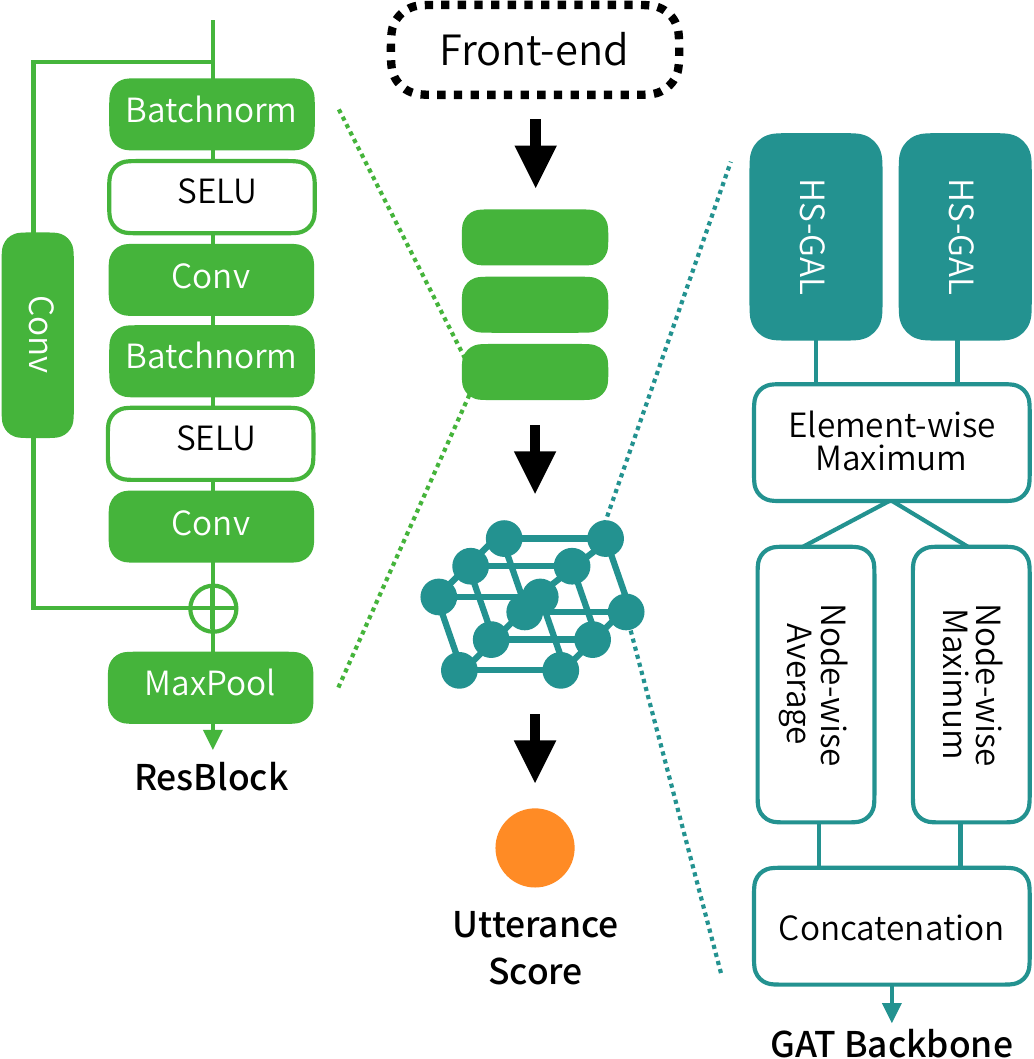}
\caption{Baseline systems architecture. We adjust the linear layer before the GAT backbone to adapt for different front-end dimensionalities. More details of HS-GAL are available in~\cite{Jung2021AASIST}.}
\label{fig:system-architecture}
\end{figure}

We have developed two baseline systems for the challenge: one that uses raw waveforms and another that employs linear frequency cepstral coefficients (LFCCs) as front-end features. The architecture of the baseline systems is shown in Figure~\ref{fig:system-architecture}.

The raw waveform system is based on the AASIST~\cite{Jung2021AASIST}, with several modifications: 
1) We reduced the number of output classes from two to one.
2) We adopted binary cross entropy with focal loss for training, setting the focusing parameter ($\gamma$) to 2.0 and the weight for positive examples ($\alpha$) to 0.25.
3) We omitted stochastic weight averaging.
4) We implemented a cosine annealing learning rate schedule with a maximum of 10 iterations and a minimum learning rate of \texttt{1e-6}.
5) We used the Adam optimizer, incorporating a weight decay of \texttt{1e-9}.

The LFCC system used 60 coefficients and 20 filters, with a 512 sample window and 160 sample hop size. The LFCC features pass through several downsampling residual convolution blocks and a linear layer connecting it to the graph attention network backend of~\cite{Jung2021AASIST}. 

We refer to the LFCC system as \textbf{B01} and the raw waveform model as \textbf{B02}. For both systems, we conducted training over 100 epochs using a fixed random seed, exclusively on the CtrSVDD training partition. We then selected the checkpoint that achieved the lowest validation EER on the CtrSVDD development partition for evaluation.
During training and evaluation, the models processed 4-second random audio segments from each utterance. Details of the implementation are available on the challenge GitHub repository\footnote{\url{https://github.com/SVDDChallenge/CtrSVDD2024_Baseline}}.

\begin{figure}[]
\centering
\includegraphics[width=0.48\textwidth]{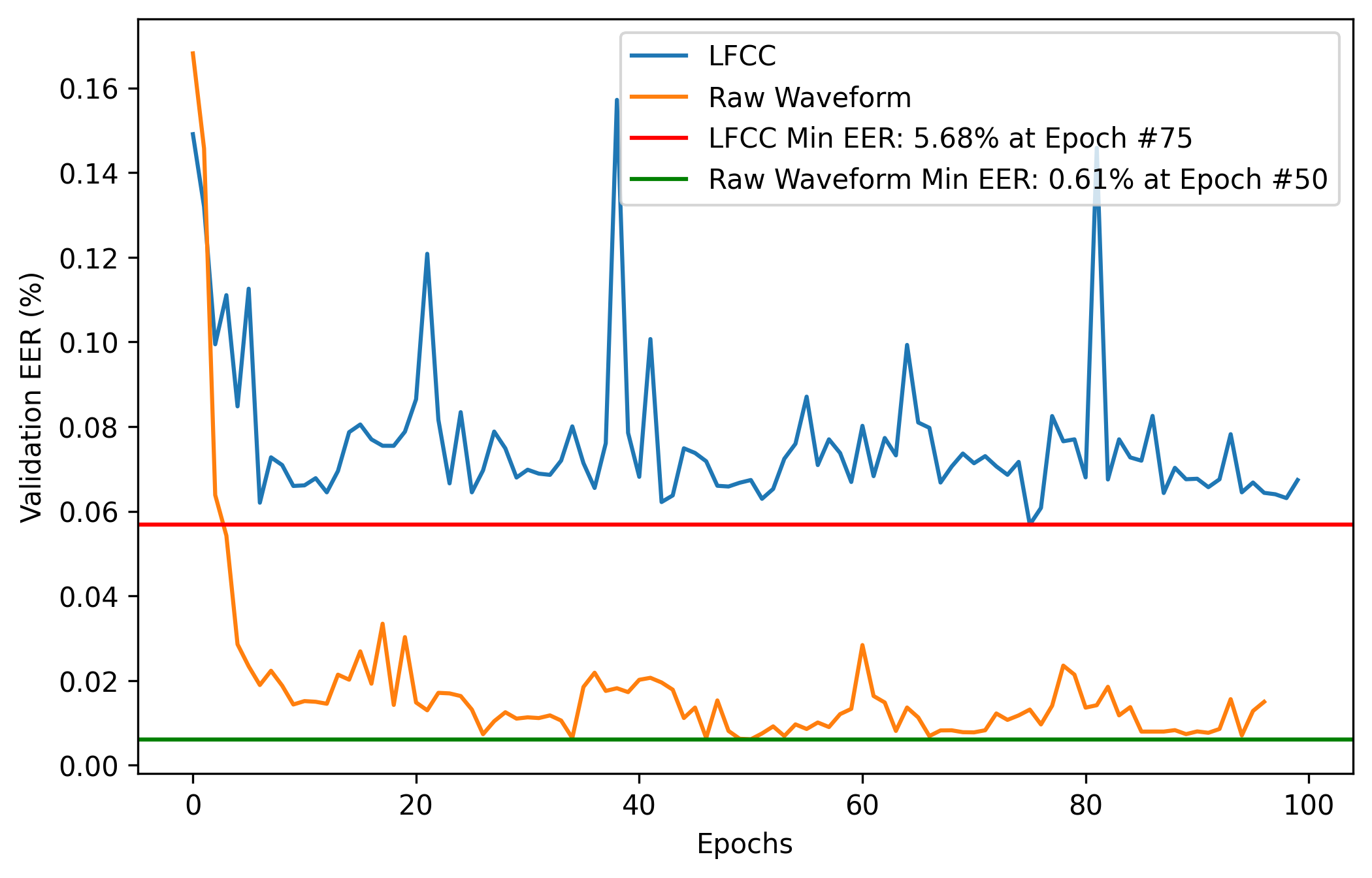}
\caption{Validation EER per training epoch. The lowest EER, indicating the checkpoint selected for evaluation, is marked by a red line for LFCC and a green line for raw waveform.  Best viewed in color.
}
\label{fig:valeer}
\end{figure}

On the CtrSVDD evaluation set, the \textbf{B01} system achieved an Equal Error Rate (EER) of 11.3697\%, while the \textbf{B02} system recorded a slightly lower EER of 10.3851\%. The performance on the validation set across each training epoch is illustrated in Figure~\ref{fig:valeer}, where we observed a rapid decline in the validation EER to below 1\%, even nearing 0. However, the performance on the evaluation set did not match this, indicating challenges in generalizing the detection of unseen singing voice deepfake generation methods. We hence encourage the participants to explore methods of improving such generalization ability.

For the WildSVDD, we will employ the same baseline systems architecture. While performance results are pending, we anticipate they will align closely with those reported in~\cite{zang23svdd}.

\section{Registration process}
Please use the following Google Form to register.

\url{https://forms.gle/FBmEYaHoVyqZSM927}

\section{Submission of results}
\label{sec: submit}

We ask the participants to submit the test set scores and system descriptions, which will be published on the challenge website. 
For reproducible research, we encourage the participants to open-source both the training code and inference code.

We have opened CodaBench\footnote{\url{https://www.codabench.org/competitions/2178}}~\cite{codalab_competitions_JMLR} for CtrSVDD results submission. 
Each team is allowed a maximum of \textcolor{red}{THREE} submissions for the entire duration of the CtrSVDD challenge for official ranking purposes. This limit is in place to ensure fairness and to encourage strategic submissions. It's important to note that CodaBench's daily submission limit is separate; our three-submission limit refers to the total allowable submissions for the challenge.

Additional submissions may be used for comparative analyses in participants' research papers. After using your THREE allotted submission opportunities, the organizers will help you register for a new CodaBench\footnote{\url{https://www.codabench.org/competitions/3004}} for further submissions. These additional submissions will be considered for research purposes ONLY and will not affect official challenge rankings. We regularly monitor submissions on CodaBench, but if you do not receive access to the new CodaBench within a day after exhausting your challenge submissions, please contact the organizers. If you wish to use the second CodaBench without participating in the challenge, please inform the organizers to gain access.



\section{Paper submission}
A special session dedicated to the SVDD challenge will be featured at SLT 2024\footnote{\url{https://2024.ieeeslt.org/}.}. Participants in the SVDD challenges may choose to submit papers via the regular submission system, which will go through SLT peer review process. 

Additionally, challenge participants have the option to submit papers describing their systems to distinct Challenge Proceedings. The challenge organizers will review these submissions. While accepted system description papers will not be indexed by IEEE, authors will be given the opportunity to showcase their work during a specific session at the workshop, facilitating a focused exchange on advancements in SVDD. 

We also plan to make all submitted system descriptions publicly available on the challenge website\footnote{\url{https://challenge.singfake.org/}}, unless participants choose not to and inform us of their decision.


\section{Important Dates}
\label{sec: dates}

Timeline of the challenge:

\begin{itemize}

\item January 8th, 2024, Release of CtrSVDD training / development data
\item January 19th, 2024, Release of the baseline system implementation for CtrSVDD
\item March 2nd, 2024, CodaBench for challenge submissions open, release of test data and baseline systems for the CtrSVDD track
\item March 29th, 2024, Release of WildSVDD dataset URLs 
\item April 2nd, 2024, Release of evaluation plan version 1.0
\item May 7th, 2024, CodaBench for research result submissions open (access upon request)
\item June 8th, 2024, SVDD Challenge Registration deadline
\item June 8th, 2024, SVDD Challenge additional training dataset permission request deadline
\item June 8th, 2024, Organizers post all available training datasets
\item June 15th, 2024, Results submission deadline (Results \& system description), CodaBench challenge submission close. Results will be publicly available on CodaBench and emailed to participants for official confirmation.
\item June 20th, 2024, SLT Paper submission
\item June 27th, 2024, SLT Paper update
\item August 30, 2024, SLT Paper notification
\item December 2nd - 5th, 2024, SVDD special session at SLT 2024
\end{itemize}

\section{Acknowledgement}
We acknowledge the contributions from ACESinger for supporting our CtrSVDD track and agree to provide participants with access to the singers' bonafide singing data. We also appreciate the support from team Opencpop \cite{wang2022opencpop}, the WeNet community, and all other bonafide data providers. 

We acknowledge Yoav Zimmerman\footnote{\url{https://www.linkedin.com/in/yoav-zimmerman-05653252/}}, Chang-Heon Han (Hanyang University, Korea), Jing Cheng (University of Rochester, USA), and Mojtaba Heydari (University of Rochester, USA) for their contributions to part of the WildSVDD data annotation.

\vfill\pagebreak
\bibliographystyle{IEEEtran}
\bibliography{mybib}

\end{document}